\documentclass[preprint2]{aastex}
\usepackage{psfig,natbib}
\shortauthors{Bower, Backer \& Sramek}
\shorttitle{Astrometric Reference Sources in the Galactic Center}
\begin{document}

\newcommand\degd{\ifmmode^{\circ}\!\!\!.\,\else$^{\circ}\!\!\!.\,$\fi}
\newcommand{\etal}{{\it et al.\ }}
\newcommand{\uv}{(u,v)}
\newcommand{\rdm}{{\rm\ rad\ m^{-2}}}

\title{VLBA Observations of Astrometric Reference Sources 
in the Galactic Center}

\author{Geoffrey C. Bower\altaffilmark{1}, 
Donald C. Backer\altaffilmark{2} \&
Richard A. Sramek\altaffilmark{1}}

\altaffiltext{1}{National Radio Astronomy Observatory, P.O. Box O, 1003 
Lopezville, Socorro, NM 87801; gbower@nrao.edu,rsramek@nrao.edu} 
\altaffiltext{2}{Astronomy Department \& Radio Astronomy Laboratory, 
University of California, Berkeley, CA 94720; dbacker@astro.berkeley.edu}

\begin{abstract}

We report here on multi-frequency VLBA observations of three extragalactic
sources within 1 degree of the Galactic Center.  These sources have been used
as astrometric reference sources for VLA and VLBA determinations of the
proper motion of Sagittarius A*, the compact nonthermal radio source
in the Galactic Center.  Each source has a main component with a
brightness temperature in excess of $10^{7.5} {\rm K}$, confirming that 
the sources are active galactic nuclei.  The sources have simple structure that
can be characterized by one or two Gaussian components.  The frequency
dependence of the structure indicates that the positions of Sgr A$^*$
determined by the VLA astrometry of 
\citet{1999ApJ...524..805B} at 4.8 and 8.4 GHz
should have an offset 
of $\sim 2$ mas.  This
offset is in the same direction as the 5 mas shift measured by 
\citet{1999ApJ...524..805B}.
The structure is unlikely to bias the 43 GHz VLBA results of
\citet{1999ApJ...524..816R}.
Motions of components in the calibrator sources could lead to
errors in the proper motion of Sgr A$^*$ on the order of 
a few km s$^{-1}$.

All three sources show frequency dependent
structure consistent with scattering which is significantly stronger than
that of the Galactic scattering model of \citet{1993ApJ...411..674T}
but significantly weaker than that of the hyperstrong Galactic Center
scattering.  Combined with other observations, this suggests the
existence of  a
new component of Galactic scattering located several kpc from the 
Galactic Center.

\end{abstract}


\section{Introduction}

The compact radio source in the Galactic Center, Sagittarius~A*,
is the best and closest candidate for a massive black hole in the
center of a galaxy \citep{1998ApJ...494L.181M}.
The source Sgr A$^*$ is
positionally coincident with a $\sim 2.6 \times 10^6 M_{\sun}$ dark
mass \citep{1997MNRAS.291..219G,1998ApJ...509..678G,2000Natur.407..349G}.
Very long baseline interferometry (VLBI) has shown that
this source has a size scale less than 1 AU and a brightness temperature in
excess of $10^9 {\rm\ K}$ 
\citep{1994ApJ...434L..59R,1998ApJ...496L..97B,1998ApJ...508L..61L,
1998A&A...335L.106K}.
For these reasons, it is inferred that Sgr~A* is a
massive black hole surrounded by a synchrotron or cyclo-synchrotron
emission region powered by
accretion 
\citep{1994ApJ...426..577M,1998ApJ...492..554N,
1993A&A...278L...1F,1998Natur.394..651M}.
Nevertheless, significant details regarding the emission region are
unknown.  In particular, we don't know whether the emission originates
in an inflow or an outflow and whether the emission region is optically
thin or thick.  These may have observational consequences.  
In particular,
some outflow models require frequency-dependent positions while most
inflow
models give frequency-independent positions.

Long-term astrometric studies
indicate that Sgr~A* shows a proper motion of $\sim 6 {\rm\ mas\ y^{-1}}$,
which is consistent with the Sun's rotation around the Galactic Center
and 
no motion of Sgr~A* with respect to the 
Galactic Center.  These results strongly confirm the hypothesis that
Sgr~A* is a massive black hole in the Galactic Center.
The Very Large Array (VLA) experiment of
\citet{1999ApJ...524..805B}
determined the position
of Sgr~A* with respect to three 
radio sources within 1 degree over 18y at 4.8 GHz and over 10y at 8.4 GHz.
\citet{1999ApJ...524..816R}
measured the position of Sgr~A* with respect to two of these same
sources over 2y with the Very Long Baseline Array (VLBA) at 43 GHz.  
Both experiments
assume that the reference sources are extragalactic and have 
frequency-independent, time-invariant positions.
In one of their highest quality epochs,
\citet{1999ApJ...524..805B}
also found a 5 mas offset between the 4.8 and 8.4 GHz
positions of Sgr~A*.  
A portion or all of this could be the result of source structure as we
will discuss below.  However, since the effect is $\sim 0.01$
times the beam size of the VLA and is $\sim 0.1$ times the scattering
size, systematic errors and refraction could contribute to the
offset.

Interstellar scattering of the
radiation along the line of sight broadens the image
of Sgr~A* and other Galactic Center sources
at radio through millimeter wavelengths 
\citep[e.g.,][]{1998ApJ...508L..61L,1994ApJ...427L..43F,
1998ApJ...505..715L}.
The scattering medium has been modeled
as a plasma at a distance of $\sim 150$ pc from the Galactic Center
with a radial extent in Galactic latitude and longitude on the order of $0.5^\circ$,
using constraints from the scattering sizes of sources physically
located in the Galactic Center, the emission measure of the plasma
and surveys for background extragalactic sources.  
The size of Sgr~A* is $\sim 1^{\prime\prime}$ at 1 GHz.
An extragalactic source behind the scattering
screen will have a size $\sim 100^{\prime\prime}$ since the scattering
of Sgr~A* is very inefficient due to its proximity to
 the scattering region.
The scattering medium, which is often referred to
as the hyperstrong scattering medium, 
requires a turbulent electron density several orders of magnitude
greater than that predicted by the best model for galactic 
electron content 
\citep{1993ApJ...411..674T}.
The model predicts that extragalactic sources in the Galactic Center
will have a size of only $\sim 0.1^{\prime\prime}$ at 1 GHz.
The model is known to be
inaccurate in the Galactic Center because it excludes 
scattered sources to avoid confusion with the hyperstrong scattering
region and because it relies on pulsar data which lack sensitivity
beyond $\sim 4{\rm\ kpc}$.

We present here VLBA observations at 2.3 GHz, 5.0 GHz and 8.4
GHz of the three astrometric reference
sources of \citet{1999ApJ...524..805B}.  In Section~2 we summarize
the observations and results.  In Section~3 we show that the sources
are very likely to be extragalactic.  In Section~4 we discuss the impact of
these observations on the VLA and VLBA astrometric results.
In Section~5 we show that the sources are scattered and discuss the
medium responsible for the scattering.  We give our conclusions
in Section~6.

\section{VLBA and VLA Observations and Results}

Observations were made with the VLBA from 2300 UT 29 August 1999  
to 0600 UT 30 August 1999.  A single
VLA antenna was also included to increase sensitivity to large-scale
structure.  Sky frequencies of 2.3, 5.0 and 8.4 GHz were observed with
total sample rates of $128 {\rm\ Mbits\ s^{-1}}$.  The 
VLA did not observe at 2.3 GHz.  
We cycled frequently through the 
sources W56 (B1742-283=J1745-2820), W109 (B1745-291=J1748-2907), 
GC 441 (B1737-294=J1740-2929) and Sgr~A*
at the three separate frequencies.    See Table~\ref{tabpos} for
source positions.  Hourly observations of NRAO~530
(J1733-1302) were also made at each frequency.  The data were
correlated in Socorro, New Mexico.  

Analysis was performed 
with the AIPS package.  A priori amplitude calibration was
applied using measured system temperatures and standard gain curves.  
High SNR fringes were detected on all 
baselines for NRAO~530.  Fringes were also found on short
baselines for the four other sources at 5.0 and 8.4 GHz.  
At 2.3 GHz fringes were detected consistently for GC 441 only.
No fringes were detected for Sgr~A* and W56 at 2.3 GHz.  An
attempt was made to reference the phase of the different sources.
This failed due to ionospheric phase fluctuations and the small
number of baselines.  Instead,
the visibility data were phase self-calibrated before imaging.
Thus, absolute position information is not recovered in these
observations.
We present the results of fitting Gaussian models to the visibility data
in Table~\ref{tabmodel}.  Zero baseline fluxes determined by 
VLA observations on 10 April 1998 (Table~\ref{tabflux}) indicate
that $90 \pm 10\%$  of the flux is recovered in the VLBA images at 5.0 and
8.4 GHz for all sources.
Monitoring of the calibrator fluxes over the past 20y shows that these sources
are slowly variable \citep{2001ApJprepbower}.
Images of GC 441 at 5.0 and 8.4 GHz are shown in Figure~1.

We also present in Table~\ref{tabflux} the results of linear polarimetric observations
at 4.8 and 8.4 GHz with the VLA of these sources and the
source J1751-253 on 10 April 1998.  
The 4.8 GHz
results were previously presented in 
\citet{1999ApJ...521..582B}.
The polarization position
angles were calibrated with observations of 3C 286.  We calculate a
rotation measure (RM) for each source.  The error in position angle is
dominated by calibration and is on the order of a few degrees.
The error in RM is on the order of $10\rdm$.  The sources are
compact in these A array observations, implying sizes
less than 300 mas.  There is some evidence in the visibilities
for diffuse 
structure around W109 but it is not clear if this is physically associated
with the source.

\section{The AGN Nature of the Reference Sources}

The brightness temperature for each source component is listed in
Table~\ref{tabmodel}.  For the brightest components in each source, $T_b \ga 10^{7.5}
{\rm\ K}$.  
Supernovae \citep[e.g.,][]{1997ApJ...486L..31M}
and galactic stellar-mass
black hole sources \citep[e.g.,][]{1995Natur.375..464H}
are the only galactic sources known to have brightness temperatures
this large.
However, each of these display light curves
inconsistent with
the roughly constant flux of the reference sources over the past 20y
\citep{2001ApJprepbower,
1998ApJ...500...51W,
1999AAS...194.4802W}.  

The sources GC 441 and W56 resemble extragalactic
jet sources with a flat spectrum core (component A) and a steep spectrum jet (component B),
while W109 resembles an unresolved flat spectrum radio core.
The spectral index for GC 441B is $\alpha=-1.2 \pm 0.4$
($S_{\nu} \propto \nu^\alpha$).  We can only find an upper limit on
$\alpha$ for W56B, which was not detected at 8.4 GHz.  
The rms in the map at the location of W56B is 0.140 mJy beam$^{-1}$.
The lack of a detection of W56B at the $5\sigma$ level implies $\alpha < -2.4 \pm 0.7$,
which indicates that this component is almost certainly
heavily resolved at 8.4 GHz.  
The fluxes of the secondary components were too low to obtain reliable fits
for their sizes in either the image or the visibility planes.

Given the absence of resemblance to known galactic sources and the
clear resemblance to extragalactic jet sources,
we conclude that the sources are extragalactic.
This strongly supports the assumptions that underly the astrometric 
conclusions of \citet{1999ApJ...524..805B}
and \citet{1999ApJ...524..816R}.

\section{The Differential Astrometric Position of Sagittarius A*}

The sources W56 and GC 441 have asymmetric structure that
is frequency dependent.  GC 441 is extended to the Northwest and
W56 is extended to the Southwest.  Both sources have greater 
asymmetry at 5.0 GHz than at 8.4 GHz.  The effect of these
asymmetries is that the astrometrically measured position of
Sgr~A* relative to these reference sources 
at 5.0 GHz will be to the East of that at 8.4 GHz.
While our VLBA results do not have absolute or relative astrometry 
between the sources, we can explore
how the reference source centroids, which are what the VLA observations
measure, vary between frequencies. Consequently we can estimate
systematic differences in the VLA differential astrometry of Sgr A$^*$.

We can estimate the 
frequency-dependent positions
by computing the centroids of the images.  In 
Table~\ref{tabfreqpos} we report the difference in the centroids of the emission
with the peak of the emission at each of the two frequencies
($\Delta\alpha_C$ and $\Delta\delta_C$ at 5.0 GHz and
$\Delta\alpha_X$ and  $\Delta\delta_X$ at 8.4 GHz).
We then give the difference in these terms,
$\Delta\alpha_{CX}=\Delta\alpha_C-\Delta\alpha_X$
and $\Delta\delta_{CX}=\Delta\delta_C-\Delta\delta_X$,
as estimates of the difference in the frequency-dependent positions.
The results
for Sgr~A* and W109 serve as estimates of the error due to
noise in the map.  However, as we discuss below, the total error
is probably dominated by resolution and opacity effects.
Combining the frequency-dependent positions
with the weighting applied by \citet{1999ApJ...524..805B}
we find that the measured position of Sgr A$^*$ at 4.8 GHz will be 
$2.0 \pm 0.1$ mas to the East and $0.6 \pm 0.4$
mas to the North of the position at 8.4 GHz.  
\citet{1999ApJ...524..805B} found in a single epoch 
that the 4.8 GHz position was $\sim 5$ mas to the East of the
8.4 GHz.  The agreement in sign and order of magnitude 
between these results suggests that we are
accounting for the dominant effect.

The discrepancy in magnitude can be explained by the difficulties of 
estimating positions with the VLA to mas-accuracy, by missing
zero-baseline flux in the VLBA results, and by offsets in the peak of emission
at the different frequencies.
The missing zero-baseline flux is almost certainly distributed
along the principal axes of the two component sources.  For example, we see
a suggestion in the 5.0 GHz image of GC 441 of an additional weak component
at a distance of 100 mas along the axis of the jet.  Since jet sources
are likely to be more extended at lower frequencies,
any additional flux will bias the results to a larger angle.
Offsets in emission peaks are also likely to play
a role at sub-mas scales since the source cores are optically
thick.  \citet{1997A&A...325..383R} have measured an offset
in the peak of emission at 13 cm and 3.6 cm of 0.7 mas
of the extragalactic radio source 1038+52A,
for example.

The differential position of Sgr~A* determined at 4.8 GHz will differ from
that at 43 GHz, as well.  This difference can be computed assuming
that all sources have single components at 43 GHz.  The expected
difference then between the \citet{1999ApJ...524..805B}
position and the
\citet{1999ApJ...524..816R}
position is 2.3 mas to the East and 0.6 mas to the North.
This is less than the error in 
the measured absolute position of Sgr A$^*$, which is
$\sim 5$ mas.

Structural variability in the sources could impact
the frequency-dependent position and measured proper motion
of Sgr A$^*$.
Assuming a proper motion of 1 mas y$^{-1}$
\citep{1994ApJ...430..467V}
for the GC 441B and W56B components, we can estimate the 
evolution of the frequency-dependent position and an
offset in the proper motion of Sgr A$^*$.
Using a weighted
average of source positions, we find a
potential bias in the measured proper motion of Sgr A$^*$ 
at 5.0 GHz of 0.1 mas y$^{-1}$ over the past 20y.  
The proper motion of Sgr A$^*$
would become more negative in right ascension and more
positive in declination, in roughly equal parts.  This brings
the measured proper motions of 
\citet{1999ApJ...524..805B} and \citet{1999ApJ...524..816R}
slightly closer together.  The differences, however,
are still greater than the $1\sigma$ errors in
the right ascension proper motion.
For both sources, the flux evolution of the A and B components
could have as substantial an impact on the centroids as their
proper motion does.

Motion of currently
undetectable components closer to the cores of the sources 
could have a much more significant impact on the proper motion
of Sgr A$^*$.  This is true for both the 4.8 and 43 GHz results.  
Additionally, many sources
at higher frequencies are known to have variable structure on
timescales of weeks to months 
\citep[e.g.,][]{1999AAS...194.6203M,1997ApJ...484..118B}.
Predicting the direction of these proper motion offsets is
not possible with these observations, either.  
First, we have observed no preferred angle for the source W109.  
Second, many sources show bent or misaligned jets between milliarcsecond
and arcsecond scales, or have misalignments between different 
frequencies \citep{1988ApJ...328..114P}.  
This implies that proper motion
measurements with accuracies better than 0.1 mas y$^{-1}$ must
be accompanied by high resolution
imaging of the astrometric reference sources.

\section{Scattering of Sources in the Galactic Center Region}

\subsection{Angular Broadening}

In the Appendix of his review Rickett has formulated the
dependence of apparent source visibility of point sources on the turbulent 
properties of the intervening medium \citep{1990ARA&A..28..561R}.
Angular scattering in the intervening medium leads to apparent normalized
visibility for a point source of the form 
\begin{equation}
\Gamma(\bar b) = e^{-{1\over 2} D(\bar b)},
\end{equation}
where $D(\bar b) = A_\nu b_x^\alpha + B_\nu b_y^\alpha$ is the phase structure 
function 
of the scattering region, and $\bar b=(b_x,b_y)$ is the vector projected 
baseline length.
This formulation allows for the general case of asymmetric turbulence with
unequal amplitudes along orthogonal, principal axes $A_\nu \ne B_\nu $; 
see \citet{1990ARA&A..28..561R} for definition of these amplitudes.
The structure function also depends quadratically on the inverse radio 
frequency
owing to the cold plasma dispersion law $A_\nu ,B_\nu \propto \nu^{-2}$.
For the extreme scattering case which is present in the Galactic Center
and with projected baselines less than the inner scale
of the turbulence, we expect $\alpha$ equals exactly 2. 
Therefore, the model predicts Gaussian 
components whose size depends quadratically on inverse frequency. 
The observed frequency dependence of the images then provides a check on 
$\alpha=2$. 

The three extragalactic sources and Sgr~A* show clear evidence
for scattering in their size as a function of frequency (Figure~2).
The sizes for Sgr~A* are consistent with those measured previously
at these and other frequencies 
\citep[e.g.,][]{ 1998ApJ...496L..97B,1998ApJ...508L..61L}.
For $\sigma_{maj} \propto \nu^{-\alpha}$,
$\alpha_{Sgr A^*}=1.90 \pm 0.04$,
$\alpha_{GC 441}=1.78 \pm 0.34$,
$\alpha_{W56}=1.98 \pm 0.14$, and
$\alpha_{W109}=1.96 \pm 0.13$.  
These are reasonably consistent with $\alpha = 2$ or slightly less. They are
marginally consistent with the Kolmogorov case for baselines greater than the
inner scale, $\alpha=1.67$.
The constant axial ratios also imply 
that the minor axes of these sources follow $\alpha=2$.  In addition, the
position angles $\phi$ are consistent with no change with
wavelength.  
Results for GC 441B and W56B are roughly consistent with the scattering but are difficult
to characterize quantitatively due to low SNR.

Additionally, fits to the visibility data at each frequency
with $\alpha$ unconstrained
consistently found $\alpha=2.0$ for the main components of each
source.  This confirms that the source
sizes found are not confused by extended structure
and that the inner scale of the Kolmogorov turbulence distribution
is greater than the longest effective
baseline on which detections were made
\citep[e.g.,][]{1994MNRAS.269...67W}.
This is on the order of $20 M\lambda$ at 8.4 GHz, which is 700 km.
In the case of Sgr A$^*$, the effective baseline
is the baseline length scaled by the ratio $f$ of the distance of the 
source to the scattering screen to the distance of the source 
to the observer.  For Sgr A$^*$, this ratio is on the order of 0.01,
implying an inner scale greater than 10 km.

\subsection{The Intermediate Strength Scattering Region}

The scattering sizes of the three extragalactic sources are two orders of
magnitude less than what is expected for an extragalactic source
behind the hyperstrong scattering region of the Galactic Center.
At 5 GHz, this size is $\sim 4^{\prime\prime}$ 
\citep{1998ApJ...505..715L}.
Thus, these sources either delineate the outer limits of the hyperstrong
Galactic Center scattering region or are viewed through holes
in the hyperstrong scattering region.  We favor the former explanation.
Each of these sources was at the center of a 
$30^\prime$-diameter field surveyed 
at 20 cm by \citet{1998ApJS..118..201L}.
The number of sources found in each field
($>12$)
was significantly in excess of the number of extragalactic sources expected
($\sim 5$)
in the absence of the hyperstrong scattering region, indicating that the
entire fields, or substantial parts, are outside of the hyperstrong 
scattering region.  Fields closer to Sgr~A*
show significantly fewer or no sources.
In order for holes in the scattering medium
to explain the weaker scattering, the scale of patchiness must be
significantly less than the field size, $30^\prime$, and the hyperstrong
scattering region must also cover a small fraction of the fields.  
\citet{1999ApJ...515..196L} estimate the scale
of patchiness at $5^\prime$ 
from the proximity of strongly scattered galactic sources to
the extragalactic source 1LC359.872+0.178, which is not scattered
by the hyperstrong scattering region.
However, the covering factor of the hyperstrong scattering region in
the field surrounding this source must be much higher than in
our fields, since this
source is the only one found in this field. 
We conclude that the
reference sources are outside of the hyperstrong scattering region.
They do not uniquely define the extent in latitude and longitude, but they do
suggest the region is bounded by $359^\circ<l<1^\circ$ and 
$|b|< 0.5^\circ$.

The scattering sizes of the reference sources
are a factor of 1.5 to 6 greater than predicted 
by the galactic scattering model of \citet{1993ApJ...411..674T}.
This is unsurprising given
the lack of sensitivity in the 
\citet{1993ApJ...411..674T}
model to the Galactic Center scattering.  These results
suggest the existence of a scattering 
region of intermediate strength between the 
\citet{1993ApJ...411..674T}
model and the hyperstrong Galactic Center scattering.

We plot in Figure~3 the
scattering sizes of our sources along with those of several extragalactic
sources and OH 1612 MHz maser spots from \citet{1992ApJ...396..686V}.  For
clarity we exclude scattered sources within $0.25^\circ$ of Sgr~A*
which are clearly associated with the hyperstrong scattering region
\citep{1994ApJ...427L..43F,1999ApJ...512..230Y}.
The extent of the region of intermediate-strength scattering is not well
constrained. 
The scattering sizes of our sources are similar to those
of some extragalactic sources in the vicinity of the Galactic Center
[e.g., J1744-312
($l=357.86^\circ,b=-1.00^\circ$), B1739-298 ($l=358.92^\circ,b=0.07^\circ$)
\citep{1998ApJS..118..201L}
and 1LC 359.872+0.178 \citep{1999ApJ...515..196L}].

Other, more weakly scattered sources provide outer bounds to the 
intermediate strength scattering region.
We can limit its extent in Galactic latitude with 
sources from
the VLBA calibrator survey \citep{1998rege.conf..155P}:
J1700-2610 ($l=356.73^\circ,b=10.02^\circ$),
J1713-2658 ($l=357.67^\circ,b=7.10^\circ$) and
J1820-2528 ($l=6.79^\circ,b=-4.94^\circ$). 
At 4cm all show flat amplitudes on
VLBA baselines as long as $250 M\lambda$, 
indicating compact component sizes less than 1 mas.
Together, these provide an upper limit of $5^\circ$ in latitude for
the intermediate-strength scattering region.

The SM is greater at negative longitudes than at positive longitudes.
For negative longitudes,
the extent in longitude of the intermediate-strength scattering
is $\ga 5^\circ$.
\citet{1992ApJ...396..686V}
found OH 1612 MHz masers with sizes of $\sim 350$ mas at negative 
longitudes as great as $6.7^\circ$.
Maser sizes for $l > +1.3^\circ$ are typically
100 mas, corresponding to 10 mas at 5 GHz.  This is supported by the lack of 
intermediate-strength scattering in J1751-253 ($l=3.73^\circ,b=0.85^\circ$).
This extragalactic object is known to be scattered (R. Hjellming,
W. Brisken, private communications) with a size at 1.6 GHz of 40 mas,
consistent with the prediction of 
\citet{1993ApJ...411..674T}.

The similar degree of scattering of the Galactic
Center OH sources and the extragalactic sources implies that the 
scattering must be distant from the Galactic Center.  This is unlike
the case of the hyperstrong scattering region which is $\sim150 {\rm\ pc}$
from the Galactic Center.

The intermediate-strength scattering region in the Galactic Center
is stronger than other strong scattering regions in the Galaxy.  
Enhanced scattering
in the Cygnus region covers $50^\circ < l < 70^\circ$ and 
produces scattering sizes of $\sim 5 {\rm\ mas}$ at 5 GHz
\citep{1991ApJ...372..132F,2000ApJsubmitdesai}.
OH 1720 MHz masers seen in W28 and W44 may also
be scattered with typical sizes of 100 mas, corresponding to
a scattering size of $\sim 10 {\rm\ mas}$ at 5 GHz 
\citep{1999ApJ...522..349C}.
OH 1612 MHz masers in the W49 region are scattered on the scale
of 100 to 200 mas \citep{1994Natur.372..754D}.
Potentially, the Galactic masers are experiencing near-field
scattering.  In this case, the strength of scattering
in these regions may equal or exceed that of the 
intermediate-strength scattering region.

Only two other individual extragalactic sources exhibit scattering as strong as we
see for the Galactic Center reference sources.  
As mentioned above, B1849+005
has a size on the order of 45 mas at 5 GHz \citep{1991ApJ...372..132F}.
Additionally,
the source NGC 6334B is very heavily scattered with a size of
$3^{\prime\prime}$ at 1.5 GHz 
\citep{1998ApJ...493..666T}.
The angular extent of the scattering region around NGC 6334B is unknown.
However, ${\rm H_20}$ masers within $2^\prime$ are unscattered.

\section{Conclusions}

We have imaged three astrometric reference sources in the Galactic
Center.  We have four principal conclusions.

1.  Based on their morphologies, brightness temperatures and
steady fluxes, we conclude that these sources are active galactic nuclei.
This supports the basic astrometric conclusions of \citet{1999ApJ...524..805B}
and \citet{1999ApJ...524..816R}.

2.  Two of the three sources are asymmetric.  The frequency 
dependence of this asymmetry accounts for the measured
frequency-dependent position of Sgr~A*.  However, we are unable
to quantitatively determine what residual frequency-dependence
there may be in the position of Sgr~A*.  Thus, we cannot
distinguish between optically thick jet models and optically
thin accretion models based on these data.  A new result requires 
denser $\uv$ coverage on baselines less than 300 km.  The proposed
A+ array of the expanded VLA will be ideal for making these
measurements.

3.  Proper motion of components in the
reference sources could contribute $0.1 {\rm\ mas\ y^{-1}}$ to
the measured proper motion of Sgr A$^*$.  This may account
for some of the discrepancy between the results of 
\citet{1999ApJ...524..805B}
and \citet{1999ApJ...524..816R}.

4.  The sources are all scattered.  All three are scattered much
less than expected from the hyperstrong Galactic Center
scattering region.  Two of the three, W56 and W109, are scattered
much more than expected by the galactic electron distribution
model of \citet{1993ApJ...411..674T}.  This suggests the existence of
an intermediate strength scattering region covering the Galactic
Center.  Several other known sources are also apparently scattered
by this region.  We infer that the region is several kpc away 
from the Galactic Center, that it covers $\ga 5^\circ$ in 
longitude and $< 5^\circ$ in latitude.  This is one of the
strongest known scattering regions in the Galaxy.  Its covering
factor, extent and relationship to the hyperstrong scattering
screen can be further
probed with observations of masers in the Galactic Center
and background extragalactic sources.

This final result points to a general deficiency in our knowledge
of scattering in the Galaxy.  The global models for electron
distribution were based on scattering observations made before
the advent of the VLBA.  The VLBA is capable of detecting and
imaging a
much greater number of sources over a broad frequency range.
This would lead to a Galactic electron model
of significantly greater accuracy and angular resolution.

\acknowledgements
The National Radio Astronomy
Observatory is a facility of the National Science Foundation operated under
cooperative agreement by Associated Universities, Inc.  
We thank Ketan Desai for assistance with fitting models to the
visibilities.

\bibliographystyle{apj}

\newpage
\plotone{f1a.ps}
\plotone{f1b.ps}
\figcaption[f1a.ps,f1b.ps]{VLBA images of GC 441 at (a) 5.0 and (b) 8.4 GHz.
The contour levels are -2, 2, 4, 8, 16, 32 and 64\% of the peak intensity
of $20.1 {\rm\ mJy~beam^{-1}}$ 
at 5.0 GHz and -1, 1, 2, 4, 8, 16, 32 and 64\% of the peak intensity
of $16.9 {\rm\ mJy~beam^{-1}}$ at 8.4 GHz.  
Image rms noise is 170 $\mu{\rm Jy~beam^{-1}}$
at 5.0 GHz and 80 $\mu{\rm Jy~beam^{-1}}$ at 8.4 GHz.  Synthetic
beams are shown in the lower left hand corners.}

\plotone{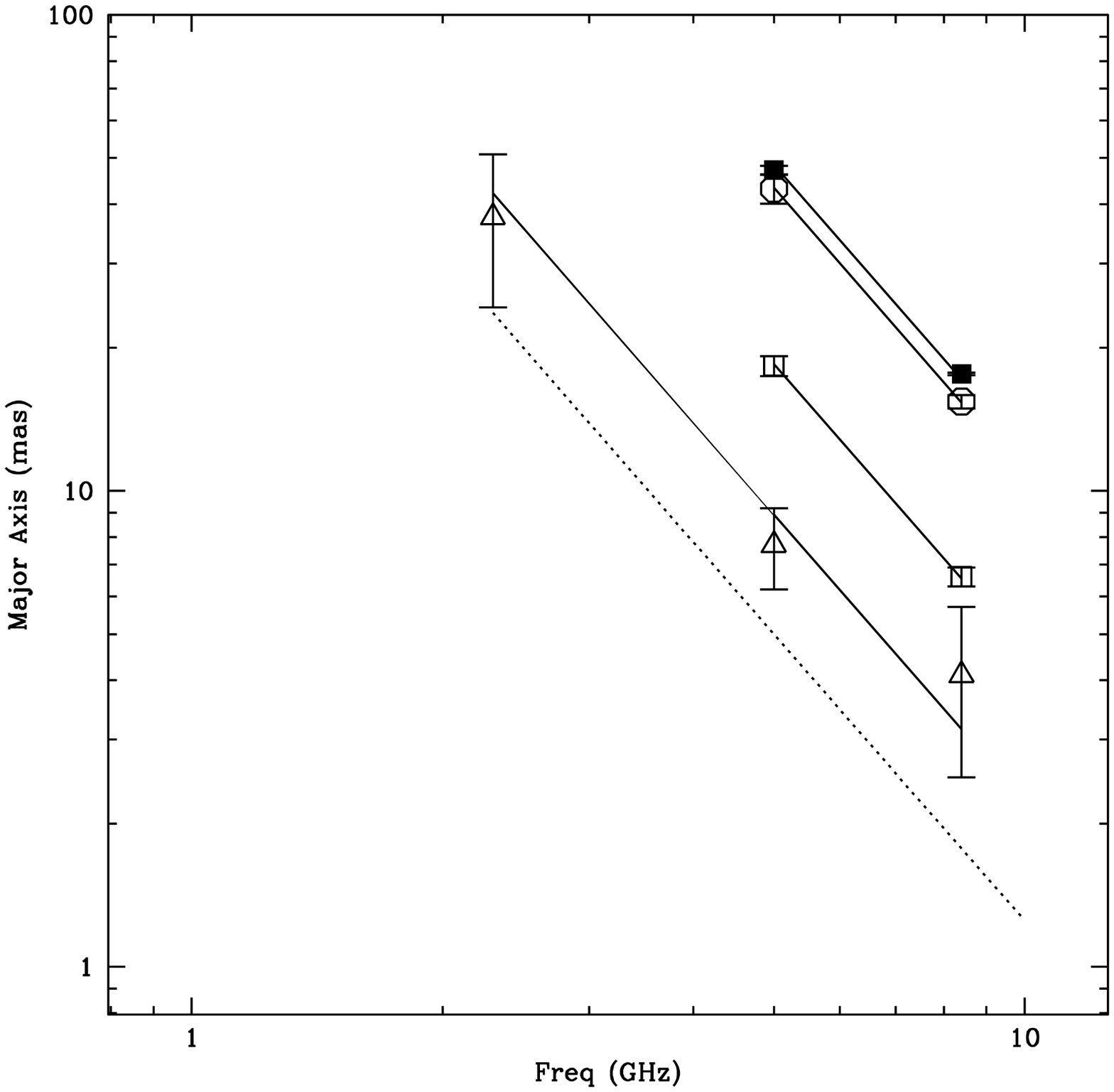}
\figcaption[f2.ps]{The major axis size of the sources plotted
as a function of frequency.  Open triangles are for GC 441A,
open squares are for W109, open octagons are for W56A and
filled squares are for Sgr~A$^*$.  The solid lines connecting the points
have a slope of -2, as expected for strong scattering.  The dotted
line is a prediction of the \citet{1993ApJ...411..674T} scattering model
towards the Galactic Center.}

\plotone{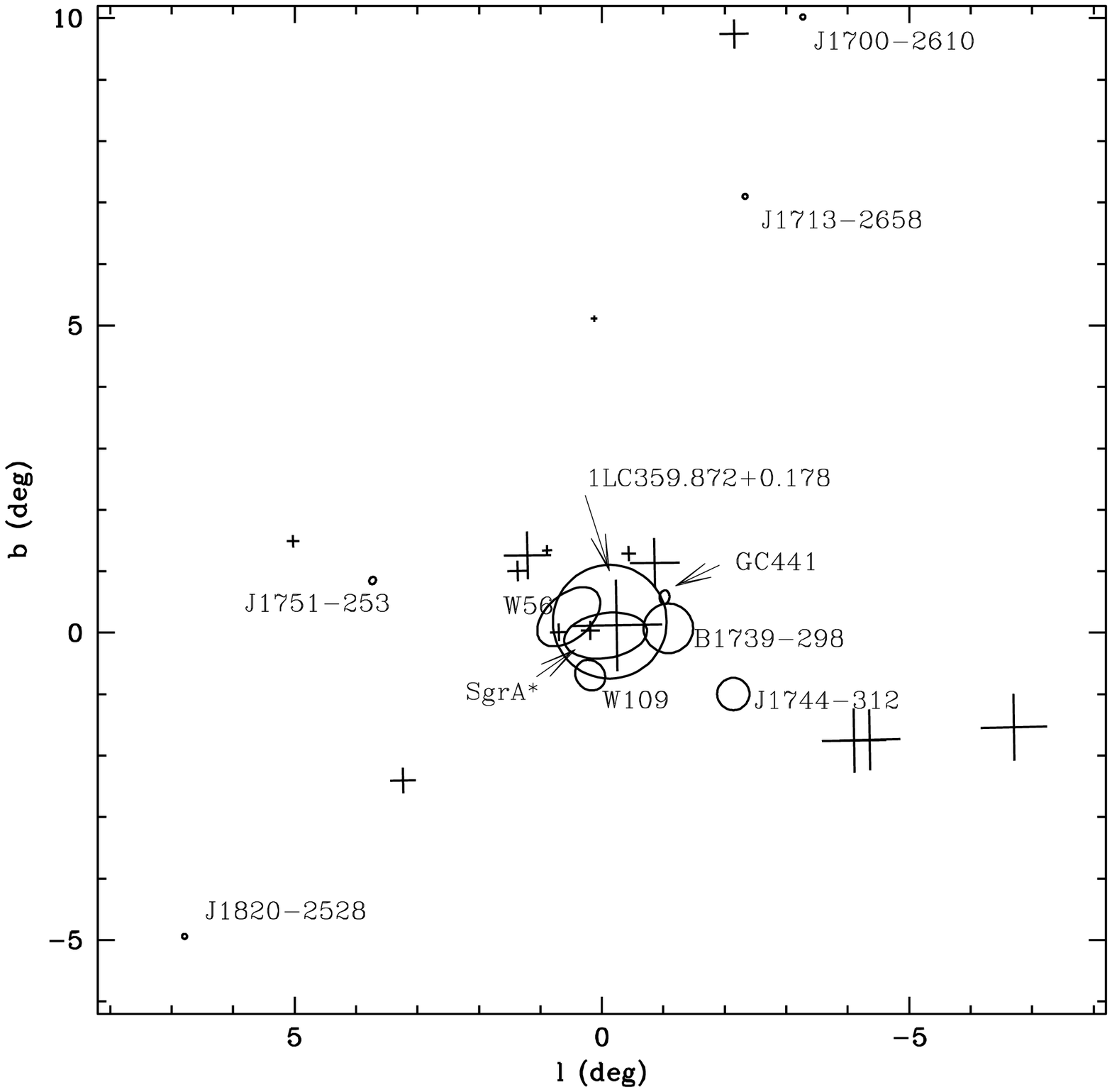}
\figcaption[f3.ps]{Scattered sources in the Galactic Center.  The size
of the  ellipse or crosshair 
indicates the size of the scattered image at 1.6 GHz with a  scaling
factor of $3^\circ {\rm\ arcsec^{-1}}$.  The program
sources are shown with their elliptical structure.  Crosshairs indicate
OH 1612 MHz masers from \citet{1992ApJ...396..686V}.  Labeled
extragalactic sources are from references in the text.}

\begin{deluxetable}{crr}
\footnotesize
\tablecaption{Source Positions (J2000) \label{tabpos}}
\tablehead{
\colhead{Source} & \colhead{RA}   & \colhead{DEC} }
\startdata
GC441   & 17:40:54.5249 & -29:29:50.290 \\
W56     & 17:45:52.4949 & -28:20:26.270 \\
W109    & 17:48:45.6841 & -29:07:39.374 \\
Sgr A*    & 17:45:40.0385 & -29:00:28.104 \\
\enddata
\end{deluxetable}

\begin{deluxetable}{lrcrrrrrrr}
\rotate
\tablewidth{23cm}
\tablecaption{Gaussian Components from VLBA Imaging \label{tabmodel}}
\tablehead{
\colhead{Source} & \colhead{$\nu$} & \colhead{Comp.} & \colhead{$\alpha$} & 
\colhead{$\delta$} & \colhead{Peak Flux} & \colhead{Major Axis} & \colhead{Axial Ratio} 
& \colhead{Position Angle} &\colhead{$\log T_b({\rm K})$} \\
                 & \colhead{(GHz)} &                 & \colhead{(mas)} & 
\colhead{(mas)} & \colhead{(mJy)} &\colhead{(mas)}      &                & 
\colhead{(deg)} & \\
}
\startdata
Sgr A$^*$   & 5.0 & A & \dots & \dots &$551.8 \pm   5.8$ 
& $ 47.2 \pm   1.0$ & $0.54 \pm 0.01$ &$ 81.2 \pm   1.1$ & 8.4 \\
         & 8.4 & A & \dots & \dots &$766.4 \pm   2.5$ 
& $ 17.6 \pm   0.1$ & $0.51 \pm 0.01$ &$ 83.4 \pm   0.7$ & 9.5 \\
GC 441   & 2.3 & A & \dots & \dots  &$ 67.7 \pm   4.6$ 
& $ 37.6 \pm  13.3$ & $0.67 \pm 0.21$ &$-17.8 \pm  32.6$ & 7.6 \\
         & 5.0 & A & \dots & \dots &$ 30.9 \pm   0.9$ 
& $  7.7 \pm   1.5$ & $0.72 \pm 0.16$ &$  4.0 \pm  11.1$ & 8.6 \\
         & 5.0 & B & $-22.0 \pm   0.4$ & $ 31.5 \pm   0.7$ &$ 10.3 \pm   0.9$ 
& $  8.2 \pm   1.2$ & \dots           & \dots            & \dots \\
         & 8.4 & A & \dots & \dots &$ 21.6 \pm   0.9$ 
& $  4.1 \pm   1.6$ & $0.54 \pm 0.23$ &$ 11.2 \pm  12.6$ & 9.1 \\
         & 8.4 & B & $-22.5 \pm   0.8$ & $ 34.1 \pm   1.5$ &$  5.5 \pm   0.9$ 
& $  9.1 \pm   2.6$ & \dots           & \dots            & \dots \\
W56      & 5.0 & A & \dots & \dots &$ 88.8 \pm   3.1$ 
& $ 42.0 \pm   3.0$ & $0.60 \pm 0.04$ &$ 48.8 \pm   4.6$ & 7.7 \\
         & 5.0 & B & $-100.7 \pm   4.4$ & $-207.3 \pm   7.1$ &$  2.4 \pm   
0.9$ & \dots & \dots & \dots & \dots \\
         & 8.4 & A & \dots & \dots &$108.4 \pm   1.6$ 
& $ 15.4 \pm   0.5$ & $0.55 \pm 0.02$ &$ 50.8 \pm   3.9$ & 8.7 \\
W109     & 5.0 & A & \dots & \dots &$ 78.8 \pm   1.7$ 
& $ 18.3 \pm   0.9$ & $0.87 \pm 0.06$ &$138.8 \pm  24.9$ & 8.2 \\
         & 8.4 & A & \dots & \dots &$ 60.2 \pm   0.9$ 
& $  6.6 \pm   0.3$ & $0.78 \pm 0.03$ &$157.9 \pm   9.1$ & 9.0 \\
\enddata
\end{deluxetable}

\begin{deluxetable}{lrrrrrrr}
\tablecaption{Polarized and Total Flux from VLA Observations \label{tabflux}}
\tablehead{
\colhead{Source}  &  \colhead{$I_{4.8}$} & \colhead{$P_{4.8}$} & 
\colhead{$\chi_{4.8}$} & \colhead{$I_{8.4}$} & \colhead{$P_{8.4}$} & 
\colhead{$\chi_{8.4}$}
& \colhead{RM} \\
                  & \colhead{(mJy)} & \colhead{(mJy)} & \colhead{(deg)} & 
\colhead{(mJy)} & \colhead{(mJy)} & \colhead{(deg)} & \colhead{(rad m$^{-2}$)} 
\\
}
\startdata
GC 441    &   44 &  $<0.09$  & \dots &   27 & $<0.5$ & \dots & \dots \\
W56      &  104 &  2.2      &  79 &   136 & 4.0 & 47 & -217 \\
W109     &   98 &  0.59     & -26 &    95 & 1.1 & 28 & 376 \\
J1751-253&  480  &  8.4      & -49 & 276 & 9.2 & -62 &  -92 \\
\enddata
\end{deluxetable}

\begin{deluxetable}{lrrrrrr}
\tablecaption{Differential Positions Between 5.0 and 8.4 GHz \label{tabfreqpos}}
\tablehead{
                 & \multicolumn{4}{c}{Centroid-Peak} & 
\multicolumn{2}{c}{Differences} \\
\cline{2-5}  \cline{6-7} \\
\colhead{Source} & \colhead{$\Delta\alpha_C$} & \colhead{$\Delta\delta_C$} 
&\colhead{$\Delta\alpha_X$} & \colhead{$\Delta\delta_X$} & 
\colhead{$\Delta\alpha_{CX}$} & \colhead{$\Delta\delta_{CX}$} \\
                 & \colhead{(mas)}            & \colhead{(mas)}            & 
\colhead{(mas)}            & \colhead{(mas)}            & \colhead{(mas)}      
      & \colhead{(mas)}             \\
}
\startdata
Sgr~A*&  -0.1 &  0.0  & -0.1 & 0.2 &  0.1 & -0.2 \\
GC 441 &  -4.0 &  6.3  & -0.9 & 1.2 & -3.2 &  5.1 \\
W56   &  -4.0 & -7.5  & -0.2 & 0.2 & -3.8 & -7.7 \\
W109  &  -0.1 &  0.3  & -0.0 & 0.0 & -0.1 &  0.4 \\
\enddata
\end{deluxetable}

\end{document}